%% file: icover.tex
\documentclass{article}

\usepackage{amsmath,amsthm, amssymb}
\usepackage{comment,xspace,enumerate,fullpage}
\usepackage[hidelinks]{hyperref}
\usepackage[capitalize]{cleveref}
\usepackage{authblk}

\usepackage{verbatim}
\usepackage{stfloats}
\usepackage{bold-extra}
\usepackage{tikz}
\usetikzlibrary{decorations.pathreplacing,calc,snakes}

\makeatletter
\newtheorem*{rep@theorem}{\rep@title}
\newcommand{\newreptheorem}[2]{%
\newenvironment{rep#1}[1]{%
 \def\rep@title{#2 \ref{##1}}%
 \begin{rep@theorem}}%
 {\end{rep@theorem}}}
\makeatother

\newtheorem{theorem}{Theorem}[section]
\newtheorem{lemma}[theorem]{Lemma}
\newtheorem{corollary}[theorem]{Corollary}
\newtheorem{observation}[theorem]{Observation}

\newcommand{\mayqed}{}

\theoremstyle{remark}
\newtheorem{remark}[theorem]{Remark}

\newtheorem{example}[theorem]{Example}
\newtheorem*{problem}{Problem}

\newreptheorem{theorem}{Theorem}
\newreptheorem{lemma}{Lemma}

  \usepackage[algo2e, noend, noline, boxed]{algorithm2e}
  \renewcommand{\CommentSty}[1]{\textnormal{#1}\unskip}
  \SetKwComment{tcm}{\{}{\}}

  \newenvironment{myalgorithm}[2][htbp]
  {%
    \setlength{\algomargin}{.2cm}
    \begin{center}
    \begin{minipage}{#2}
    \begin{algorithm2e}[#1]
    \small
     \let\Par=\par
       \def\par{\endgraf\vspace{.1cm}}
           \SetKw{To}{to}%
       \SetKw{Downto}{downto}%
           \SetKw{Or}{or}%
       \SetKwFor{Algo}{Algorithm}{}{}%
      \vspace{.15cm}%
   }
   {%
     \let\par=\Par\end{algorithm2e}%
     \end{minipage}%
     \end{center}%
   }

  \newcommand\twocol[2]{%
  \begin{center}%
  \begin{minipage}[t]{0.55\textwidth}%
  \vspace{0pt}%
  {#1}%
  \end{minipage}\hfill%
  \begin{minipage}[t]{0.45\textwidth}%
  \vspace{0pt}%
  {#2}%
  \end{minipage}%
  \end{center}}

  \newcommand{\floor}[1]{\left\lfloor #1 \right\rfloor}
  
  \newcommand{\Oh}{\mathcal{O}}

  \DeclareMathOperator{\lcp}{lcp}
  
  \newcommand{\C}{\mathcal{C}}
  \newcommand{\A}{\mathcal{A}}

  \DeclareMathOperator{\maxgap}{maxgap}
  \DeclareMathOperator{\dist}{dist}

  \renewcommand{\P}{\mathcal{P}}
  
  \newcommand{\sub}{\subseteq}
  \newcommand{\sm}{\setminus}

  \newcommand{\Occ}{\mathit{Occ}}
  \newcommand{\SolidOcc}{\mathit{SolidOcc}}
  \newcommand{\NonSolidOcc}{\mathit{NonSolidOcc}}
  \newcommand{\eps}{\varepsilon}

  \newcommand{\ShortestCover}{{\sl ShortestCover}}
  \newcommand{\TestCover}{{\sl TestCover}}

  \title{Covering Problems for Partial Words \\ and for Indeterminate Strings\footnote{%
  A preliminary version of this article appeared as~\cite{DBLP:conf/isaac/CrochemoreIKRRW14}.\newline
  Tomasz Kociumaka is supported by Polish budget funds for science in 2013-2017 as a research project under the 'Diamond Grant' program.
   Jakub Radoszewski receives financial support of Foundation for Polish Science.}}

\author[1,2]{Maxime Crochemore}
\author[1,3]{Costas S.\ Iliopoulos}
\author[4]{Tomasz Kociumaka}
\author[4]{\mbox{Jakub Radoszewski}}
\author[4,5]{Wojciech Rytter}
\author[4]{Tomasz Wale\'n}

\affil[1]{Department~of Informatics, King's College London, UK}
\affil[ ]{\url{[maxime.crochemore,c.iliopoulos]@kcl.ac.uk}}
\affil[2]{Universit\'e Paris-Est, France}
\affil[3]{Faculty of Engineering, Computing and Mathematics, University of Western Australia, Perth, Australia}
\affil[4]{Faculty~of Mathematics, Informatics and Mechanics, University of Warsaw, Poland}
\affil[ ]{\url{[kociumaka,jrad,rytter,walen]@mimuw.edu.pl}}
\affil[5]{Faculty of Mathematics and Computer Science,     Copernicus University, Toru\'n, Poland}

  \date{\vspace{-5ex}}

\begin{document}
  \maketitle
  \begin{abstract}
    We consider the problem of computing a shortest solid cover of an indeterminate string.
    An indeterminate string may contain non-solid symbols, each of which specifies a subset
    of the alphabet that could be present at the corresponding position.
    We also consider covering partial words, which are a special case of indeterminate strings
    where each non-solid symbol is a don't care symbol.
    We prove that indeterminate string covering problem and partial word covering problem are NP-complete
    for binary alphabet and show that both problems are fixed-parameter tractable with respect to $k$,
    the number of non-solid symbols.
    For the indeterminate string covering problem we obtain a $2^{\Oh(k\log k)} + n k^{\Oh(1)}$-time algorithm.
    For the partial word covering problem we obtain a $2^{\Oh(\sqrt{k}\log k)} + nk^{\Oh(1)}$-time algorithm.
    We prove that, unless the Exponential Time Hypothesis is false, no $2^{o(\sqrt{k})} n^{\Oh(1)}$-time
    solution exists for either problem, which shows that our algorithm for this case is close to optimal.
    We also present an algorithm for both problems which is feasible in practice.
  \end{abstract}

  \section{Introduction}
  A classic string is a sequence of symbols from a given alphabet $\Sigma$.
  In an \emph{indeterminate} string, some positions may contain, instead of a single symbol from $\Sigma$
  (called a \emph{solid} symbol), a subset of $\Sigma$.
  Such a \emph{non-solid} symbol can be interpreted as information that the exact symbol at the given position
  is not known, but is suspected to be one of the specified symbols.
  The simplest type of indeterminate strings are \emph{partial words}, in which every non-solid symbol
  is a don't care symbol, denoted here $\diamondsuit$ (other popular notation is $*$).

  Motivations for indeterminate strings can be found in computational biology,
  musicology and other areas.
  In computational biology, analogous juxtapositions may count as matches in protein sequences.
  In fact the FASTA format\footnote{\url{http://en.wikipedia.org/wiki/FASTA\_format}}
  representing nucleotide or peptide sequences specifically includes indeterminate letters.
  In music, single notes may match chords, or notes separated by an octave may match;
  see \cite{DBLP:journals/jda/HolubSW08}.

  Algorithmic study of indeterminate strings is mainly devoted to pattern matching.
  The first efficient algorithm was proposed by Fischer and Paterson for strings with don't care
  symbols \cite{FischerPaterson}.
  Faster algorithms for this case were afterwards given
  in \cite{DBLP:conf/stoc/MuthukrishnanP94,DBLP:conf/focs/Indyk98a,DBLP:conf/soda/Kalai02a}.
  Pattern matching for general indeterminate strings, known as generalized string matching,
  was first considered by Abrahamson~\cite{DBLP:journals/siamcomp/Abrahamson87}.
  Since then numerous variants of pattern matching in indeterminate strings were considered.
  There were also practical approaches to the original problem;
  see \cite{DBLP:journals/jda/HolubSW08,DBLP:journals/ijfcs/SmythW09} for some recent examples.
  A survey on partial words, related mostly to their combinatorics,
  can be found in a book by Blanchet-Sadri~\cite{PartialWords}.

  The notion of cover belongs to the area of quasiperiodicity, that is, a generalization
  of periodicity in which the occurrences of the period may overlap \cite{Apo93}.
  A \emph{cover} of a classical string $s$ is a string that covers all positions of $s$ with its occurrences.
  Covers in classical strings were already extensively studied.
  A linear-time algorithm finding the shortest cover of a string was given
  by Apostolico et al.\ \cite{DBLP:journals/ipl/ApostolicoFI91} and later on improved
  into an on-line algorithm by Breslauer \cite{DBLP:journals/ipl/Breslauer92}.
  A linear-time algorithm computing all the covers of a string
  was proposed by Moore \& Smyth \cite{DBLP:conf/soda/MooreS94}.
  Afterwards an on-line algorithm for the all-covers problem was given by
  Li \& Smyth \cite{DBLP:journals/algorithmica/LiS02}.
  Other types of quasiperiodicities are seeds \cite{DBLP:journals/algorithmica/IliopoulosMP96,DBLP:conf/soda/KociumakaKRRW12}
  and numerous variants of covers and seeds, including approximate and partial covers and seeds.

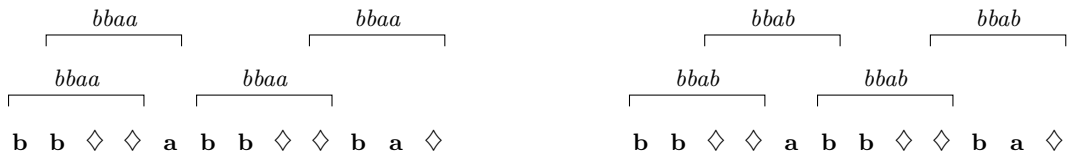
\begin{figure}[h]
      \begin{center}
    \twocol{
      \input{_fig1}
    }{
      \input{_fig2}
    }
   \end{center}
  
    \caption{
      Partial word $bb\diamondsuit\diamondsuit abb\diamondsuit\diamondsuit ba\diamondsuit$
      with its two shortest covers.
      Note that the same non-solid symbol can match two different solid symbols for two different occurrences
      of the same cover.
    }
    \label{fig:example}
  \end{figure}

  \textbf{The main problem} considered here is as follows:
  Given an indeterminate string, find the length of its shortest solid cover; see Figure~\ref{fig:example}.
  We can actually compute a shortest solid cover itself
  and all the lengths of solid covers, at no additional cost in the complexity.
  However, for simplicity we omit the description of such extensions in this version of the paper.
  
  Throughout the paper we use the following notations:
  $n$ for the length of the given indeterminate string,
  $k$ for the number of non-solid symbols in the input,
  and $\sigma$ for the size of the alphabet.
  We assume that $2 \le \sigma \le n$ and that each non-solid symbol in the indeterminate string is represented by
  a bit vector of size $\sigma$.
  Thus the size of the input is $\Oh(n+\sigma k)$.

  The first attempts to the problem of indeterminate string covering were made in
  \cite{DBLP:conf/stringology/AntoniouCIJL08,DBLP:conf/stringology/BariRS09,DBLP:journals/njc/IliopoulosMMPST03}.
  However, they considered indeterminate strings as covers and presented some partial results for this case.
  The common assumption of these papers is that $\sigma = \Oh(1)$;
  moreover, in \cite{DBLP:conf/stringology/AntoniouCIJL08,DBLP:conf/stringology/BariRS09}
  the authors considered only so-called conservative indeterminate strings, for which $k = \Oh(1)$.

  \medskip
  \noindent
  \textbf{Our results:}
  In Section~\ref{sec:k-sigma} we show an $\Oh(n\sigma^{k/2}k)$-time algorithm
  for covering indeterminate strings with a simple implementation.
  In Section~\ref{sec:k} we obtain an $2^{\Oh(k\log k)} + nk^{\Oh(1)}$-time algorithm.
  In the same section we devise a more efficient solution for partial words with
  $2^{\Oh(\sqrt{k}\log k)} + nk^{\Oh(1)}$-time complexity.
  Finally in Section~\ref{sec:NP-hard} we show that both problems are NP-complete
  already for binary alphabet.
  As a by-product we obtain that under the Exponential Time Hypothesis
  no $2^{o(\sqrt{k})} n^{\Oh(1)}$-time solution exists for both problems.

  \section{Preliminaries}
  An \emph{indeterminate string} (\emph{i-string}, for short) $T$ of length $|T|=n$ over a finite alphabet $\Sigma$
  is a sequence $T[1]\ldots T[n]$ such that $T[i] \subseteq \Sigma$, $T[i] \ne \emptyset$.
  If $|T[i]|=1$, that is, $T[i]$ represents a single symbol of $\Sigma$, we say that
  $T[i]$ is a \emph{solid} symbol.
  For convenience we often write that $T[i]=c$ instead of $T[i]=\{c\}$
  in this case ($c \in \Sigma$).
  Otherwise we say that $T[i]$ is a \emph{non-solid} symbol.
  In what follows, by $k$ we denote the number of non-solid symbols in the considered i-string $T$
  and by $\sigma$ we denote $|\Sigma|$.
  If $k=0$, we call $T$ a (solid) string.
  We say that two i-strings $U$ and $V$ \emph{match} (denoted as $U \approx V$)
  if $|U|=|V|$ and for each $i=1,\ldots,|U|$ we have $U[i] \cap V[i] \ne \emptyset$.

  \begin{example}
    Let $A=a\,\{b,c\}$, $B=a\,\{a,b\}$, $C=aa$ be indeterminate strings
    ($C$ is a solid string).
    Then $A\approx B$ and $B\approx C$ but $A\not\approx C$.
  \end{example}

  If all $T[i]$ are either solid or equal to $\Sigma$,
  then $T$ is called a \emph{partial word}.
  In this case, the non-solid ``don't care'' symbol is denoted as $\diamondsuit$.

  By $T[i..j]$ we denote a factor $T[i]\ldots T[j]$ of $T$.
  If $i=1$,  the factor is called a prefix and if $j=n$, it is called a suffix of $T$.
  We say that a pattern i-string $S$ occurs in a text i-string $T$ at position $j$ if $S$ matches $T[j..j+|S|-1]$.
  We define the \emph{occurrence set} of $S$ in $T$, denoted $\Occ(S,T)$, as the set of all such positions~$j$.
  We say that $S$ is a \emph{solid prefix} of $T$ if $S$ is a solid string that matches the prefix $T[1..|S|]$.
  
  A \emph{cover} of $T$ is a solid string $S$ such that each position $i$ of $T$
  is covered by an occurrence of $S$ in $T$, i.e., $\Occ(S,T)\cap\{i-|S|+1,\ldots,i\}\ne \emptyset$.
  If $S$ is a cover of $T$, any subset $\C\sub \Occ(S,T)$ already satisfying the latter property for all $i=1,\ldots,n$
  is called a \emph{covering set} of $S$.
   
  \begin{observation}\label{obs:atmost2}
    Let $\C$ be a minimal covering set of a cover $S$ of  $T$.
    Then each position of $T$ is covered by one or two occurrences $T[i..i+|S|-1]$ for $i\in \C$.
  \end{observation}

  \begin{remark}
    The shortest cover of an i-string $T$ need not be one of the shortest covers
    of the solid strings matching $T$.
    E.g., for a partial word $T=a \diamondsuit b$ over $\Sigma=\{a,b\}$,
    the shortest cover $ab$ has length 2, whereas neither of the solid strings $aab$, $abb$ has
    a cover of length 2.
  \end{remark}
  
  \subsection{Algorithmic Tools}

  For convenience, we compute the set $T[i] \cap T[j]$ for each pair $T[i]$, $T[j]$ of non-solid symbols of $T$,
  and label different such sets with different integers,
  so that afterwards we can refer to any of them in $\Oh(1)$ space.
  In particular, after such $\Oh(\sigma k^2)$-time preprocessing, we can check in $\Oh(1)$ time if any
  two positions of $T$ match.

  A longest common prefix (LCP) query in $T$, denoted as $\lcp(i,j)$, is a query for the length of the
  longest matching prefix of the i-strings $T[i..n]$ and $T[j..n]$.
  Recall that for a solid string we can construct in $\Oh(n)$ time a data structure that
  answers LCP-queries in $\Oh(1)$ time, see \cite{AlgorithmsOnStrings}.
  In the following lemma we note that an LCP-query in an i-string can be reduced to $\Oh(k)$ LCP-queries in a solid string.

  \begin{lemma}\label{lem:LCP}
    For an i-string with $k$ non-solid symbols, after $\Oh(nk^2)$-time preprocessing, one can compute the length of the longest common prefix
    of any two suffixes of $T$ in $\Oh(k)$ time.
  \end{lemma}
    \begin{proof}
    For an i-string $T$, by $T_{\$}$ we denote a solid string obtained by substituting
    respective non-solid symbols in $T$ by $ \$_1,\ldots,\$_k \notin \Sigma$.
    To answer an LCP-query in $T$, we repetitively ask LCP-queries in $T_\$$, treating
    non-solid symbols specially; see the following pseudocode.

    \begin{center}
      \begin{myalgorithm}[H]{8cm}
      \Algo{$\lcp(i,j)$}{
        $\mathit{res}:=0$\;
        \While{$i \le n$ $\mathbf{and}$ $j \le n$ $\mathbf{and}$ $T[i] \approx T[j]$}{
          $p:=\max(1,\lcp_{T_\$}(i,j))$\;
          $i:=i+p$; $j:=j+p$; $\mathit{res}:=\mathit{res}+p$\;
        }
        \Return{$\mathit{res}$}\;
      }
      \end{myalgorithm}
    \end{center}
  
    We obtain $\Oh(k)$ query time after additional $\Oh(\sigma k^2) = \Oh(nk^2)$-time preprocessing
    required for checking if a given pair of symbols in $T$ match.
  \mayqed\end{proof}

  Lemma~\ref{lem:LCP} lets us efficiently check if given pairs of factors of
  an i-string match and thus it has useful consequences.

  \begin{corollary}\label{cor:occ}
    Given i-strings $S$ and $T$ of total length $n$ containing $k$ non-solid symbols in total,
    one can compute $\Occ(S,T)$ in $\Oh(nk^2)$ time.
  \end{corollary}

  \section{Simple Algorithm Parameterized by $k$ and $\sigma$}\label{sec:k-sigma}
    Note that a solid string of length at least $\frac{n}{2}$ is a cover of $T$ if and only if 
    it occurs both as a prefix and as a suffix of $T$. 
    In other words, $T$ has a cover of length $m\ge \frac{n}{2}$ if and only if $\lcp(1,n-m+1)=m$.
    Therefore, Lemma~\ref{lem:LCP}  lets us easily solve the covering problem for cover lengths
  at least half of the word length.
  In this section we search only for the covers of length at most $\floor{\frac{n}{2}}$.
  
  Let $T$ be an i-string of length $n$ with $k$ non-solid symbols.
  We assume that $T[1..\lfloor\frac{n}{2}\rfloor]$ contains at most $\frac{k}{2}$ non-solid symbols; otherwise we reverse the i-string.

  For an increasing list of integers $L\;=\; [i_1,i_2,i_3,\ldots,i_m]$, $m \ge 2$, we define
  $$\maxgap(L)\,=\, \max\{i_{t+1}-i_t\;:\; t=1,\ldots,m-1\}.$$
  This notion lets us characterize covering sets:
  
 \begin{observation}\label{obs:maxgap}
  A set $\P \subseteq \Occ(S,T)$ is a covering set for $S$
  if $1 \in \P$ and $\maxgap(\P \cup \{n+1\}) \le |S|$.
  \end{observation}

  We introduce a \ShortestCover$(S,L)$ subroutine which, for a given solid prefix $S$ of $T$
  and an increasing list of positions $L$, checks if there is a cover of $T$ which is a prefix of $S$
  and admits a covering set $\C \sub L$. If so, the procedure returns the length of the shortest
  such cover.
  In this section we only use this subroutine for $L=\{1,\ldots,n\}$.
  
  A pseudocode can be found below.
  Correctness of the algorithm follows from the fact that
  $$\mbox{\ShortestCover}(S,L) \,=\, \min\Big\{j\,:\,\maxgap\Big(\bigcup_{t \ge j} L_t \cup \{n+1\}\Big) \le j\Big\},$$
  where $L_j = \{i \in L\,:\, \lcp(S,T[i..n]) = j\}$.
  
  \begin{center}
    \begin{myalgorithm}[H]{12cm}
    \Algo{\ShortestCover$(S,L)$}{
      \KwIn{$S$: a solid prefix of $T$; $L$: a sublist of $\{1,\ldots,n\}$}
      \KwOut{The length of the shortest cover which is a prefix of $S$ and
      has a covering set being a sublist of $L$}
      \hspace*{-0.6cm}{\bf preprocessing:\\ }
      \lForEach{$i \in L$}{$\dist[i]:=\lcp(S,T[i..n])$
      }
      $D\;:=\;\{\;\dist[i]\,:\,i \in L\;\}$\;
      \lForEach{$j \in D$}{$L_j\;:=\; \{\,i \in L\,:\;\dist[i]=j\,\}$
      }
      $L:=L \cup \{n+1\}$\;
      \hspace*{-0.6cm}{\bf processing:\\ }
      \ForEach{$j \in D$ in increasing order}{
        \lIf{$\maxgap(L) \le j$}{\Return{$\maxgap(L)$}
        }
        \lForEach{$i \in L_j$}{remove $i$ from $L$
        }
      }
      \Return{no solution\;}
    }
    \end{myalgorithm}
  \end{center}
  
    \begin{lemma}\label{lem:ShortestCover1}
    The algorithm \ShortestCover$(S,L)$ works in $\Oh(nk)$ time
    assuming that the data structure of Lemma~\ref{lem:LCP} is accessible.
  \end{lemma}
  \begin{proof}
    Assume that we update $\maxgap(L)$ each time we remove an element from the list.
    Then $\maxgap(L)$ may only increase.
    Each operation on the list $L$, including update of $\maxgap(L)$, is performed in $\Oh(1)$ time.

    By Lemma~\ref{lem:LCP}, all $\lcp$ values can be computed in $\Oh(nk)$ time.
    The lists $L_j$ can be easily computed in total time $\Oh(n)$.
  \mayqed\end{proof}

  Any cover of $T$ is a solid prefix of $T$, so a cover of length at most $\floor{n/2}$
  is a prefix of a solid prefix of $T$ of length $\floor{n/2}$.
  By the assumption made in the beginning of this section, $T$ has at most $\sigma^{k/2}$
  solid prefixes of length $\floor{n/2}$.
  For each of them we run the \ShortestCover$(S,L)$ algorithm with $L=\{1,\ldots,n\}$.
  Lemma~\ref{lem:ShortestCover1} implies the following result.

  \begin{theorem}
    The shortest cover of an i-string with $k$ non-solid symbols can be computed
    in $\Oh(n\sigma^{k/2}k)$ time.
  \end{theorem}

  \section{Algorithm Parameterized by $k$}\label{sec:k}
  For an i-string $U$ of length $m$ and a position $i \in \Occ(U,T)$, we define:
  $$U \odot i = U[1] \cap T[i],\ldots,U[m] \cap T[i+m-1].$$
  \begin{example}
    Let $T=bb\diamondsuit\diamondsuit abb\diamondsuit\diamondsuit baa$ and $U=b \diamondsuit a \diamondsuit$.
    Then

    \smallskip
    \centerline{
      $U \odot 1 = U \odot 6 = bba \diamondsuit$,
      $U \odot 2 = b \diamondsuit aa$,
      $U\odot 3 = U \odot 7 = b \diamondsuit ab$,
      and $U \odot 9 = bbaa$.
    }
  \end{example}

  \noindent
  If $U\odot i$ is a solid string, we call an occurrence of $U$ at position $i$ \emph{solid},
  and \emph{non-solid} otherwise. By $\SolidOcc(U,T)$ we denote the list of all
  solid occurrences of $U$ in $T$, and by $\NonSolidOcc(U,T)$ --- the list of all non-solid occurrences.
  We say that $S$ is a \emph{$\odot$-prefix} of $T$ if $S$ is a solid string such that $S=T[1..|S|]\odot i$ for some position $i$.
  Note that every $\odot$-prefix of $T$ is a solid prefix of $T$. However, a $\odot$-prefix can be specified in $\Oh(1)$ space by $|S|$ and $i$.

  A position $i$ is called \emph{ambiguous} if $T[1+\ell]$ and $T[i+\ell]$ are both non-solid for some integer $\ell$.
  The set of ambiguous positions in $T$ is denoted as $\A$.
  Note that $|\A|\le k^2$.
  The following simple observation is an important tool in our algorithms.

  \begin{observation}\label{obs:main}
  Let $U$ be a prefix of $T$.  If $U$ has a non-solid occurrence at position $i$, then $i$ is an ambiguous position.
  \end{observation}
  
  We classify the solid covers of $T$ into those which are  $\odot$-prefixes of $T$ and those which are not.
  Note that each $\odot$-prefix of $T$ is uniquely determined by its length and the position $i$,
  and thus there are $\Oh(n^2)$ $\odot$-prefixes of $T$. Consequently,
  it is straightforward to devise an $\Oh(n^3k)$-time algorithm 
  checking which of them are covers. Below we present a more efficient solution, which takes $\Oh(nk^4)$ time.
  Detecting covers which are not $\odot$-prefixes is
  more difficult; as we show in Section~\ref{sec:NP-hard}, the whole problem is NP-hard.
  
  \subsection{Covering with $\odot$-Prefixes}
The following result is a technical generalization of Lemma~\ref{lem:ShortestCover1}. 

  \begin{lemma}\label{lem:collection}
    Let $\C$ be a collection of pairs $(S,L)$, where each $S$ is a $\odot$-prefix of $T$
    and $L\subseteq \{1,\ldots,n\}$ contains some positions of $T$.
    If $|\C| \le n$ and $\sum_{(S,L) \in \C} |L| = \Oh(nk^2)$ then
    \ShortestCover$(S,L)$ for all instances $(S,L) \in \C$ can be computed in $\Oh(nk^3)$ time.
  \end{lemma}
  \begin{proof}
    First, let us focus on the processing phase of the $\mbox{\ShortestCover}(S,L)$ algorithm.
    Suppose we have already computed the set $D$ (represented as an increasing list)
    and the lists $L_0,\ldots,L_n$ (stored in a table with \textit{null} entries for $i \not\in D$),
    and that we store a pointer to the position of $x$ in $L$ together with every $x\in L_i$.
    Then, the processing phase works in $\Oh(|L|)$ time since 
    $\maxgap$ of the list can be updated in constant time upon deletion of its elements.
    This gives $\Oh(nk^2)$ time across all instances.

    We perform the preprocessing phase of $\mbox{\ShortestCover}(S,L)$ for all $(S,L) \in \C$ simultaneously.
    The first part is computation of $\dist$ values.
    For all $i \in L$ we first compute
    $$\lcp(T[1..|S|],T[i..n])$$
    using LCP-queries for $T$ (Lemma~\ref{lem:LCP}).
    Afterwards, for all $\Oh(k)$ non-solid positions in $T[1..|S|]$ we check if the corresponding solid symbol in $S$ matches the respective position in $T[i..n]$.
    This takes $\Oh(|L| k)$ time per instance, which yields $\Oh(nk^3)$ time in total.
    After all $\dist$ values have been computed, we construct the sets $D$ for all instances at once using
    bucket sort in $\Oh(nk^2)$ time.
    
    Then we process instances consecutively.
    We use a global table of size $n+1$ to store (pointers to) the lists $L_0,\ldots,L_n$,
    so that we can access any of these lists in constant time.
    This allows to construct the lists in $\Oh(|L|)$ time for a given instance.
    In the same time complexity we also clean the table after processing the instance.
    This gives $\Oh(nk^2)$ time across all instances.
  \mayqed\end{proof}

  \begin{theorem}\label{thm:solid}
    The shortest cover among all $\odot$-prefixes can be computed in $\Oh(nk^4)$ time.
  \end{theorem}
  \begin{proof}
    We need to find a pair $(m,i)$ with $m$ smallest possible such that $S=T[1..m]\odot i$ is a $\odot$-prefix
    which covers the i-string $T$.
    
    The algorithm checks all the $\Oh(k)$ possibilities for the number of non-solid symbols in $T[1..m]$. In what follows, we assume
    that this value is fixed, which restricts $m$ to some interval $[b,e]$
    such that $T[b+1..e]$ is solid.
    
    Let $U=T[1..b]$. We apply Corollary~\ref{cor:occ} to
    compute $E=\SolidOcc(U,T)$ and $H=\NonSolidOcc(U,T)$ in $\Oh(nk^2)$ time.
    The positions $j\in E$ of solid occurrences are naturally partitioned according to the value of $U\odot j$.
    This partitioning can be implemented in $\Oh(nk)$ time using radix sort, because strings $U\odot j$ may differ only at $\Oh(k)$ positions
    corresponding to non-solid symbols in $U$.
    Next, using Lemma~\ref{lem:LCP}, for each partition class $P$ we determine a \emph{representative} $r_P$, which maximizes $\ell_j:=\lcp(T[1..e],T[j..n])$
    among $j\in P$.
    
    Recall that the sought value of $i$ satisfies $i\in \SolidOcc(U,T)$. Observe that $S$ is a prefix of $T[1..\ell_{r_P}]\odot r_P$ for the class $P\sub E$ containing $i$.
    Moreover, if $S$ also occurs at some position $j$, then $j\in P$ or $j\in H$.
    Thus, $S$ can be detected by the \ShortestCover\ procedure applied for each partition class
    $P$ to $(T[1..\ell_{r_P}]\odot r_P, P\cup H)$.
    We check all suitable cases using Lemma~\ref{lem:collection}. Note that $\sum_P (|P\cup H|)\le |\SolidOcc(U,T)|(1+|H|)=\Oh(nk^2)$,
    since $H\sub \A$ (by \cref{obs:main}) and $|\A|\le k^2$.
        The time complexity is $\Oh(nk^3)$, which needs to be multiplied by the $\Oh(k)$
    choices we have made in the first step of the algorithm. 
  \mayqed\end{proof}
  
   \subsubsection*{Example}
  Consider the i-string $T=bb \diamondsuit abb \diamondsuit ab  b \diamondsuit babbb \diamondsuit \diamondsuit$
  of length $18$.
  We divide the positions in $T$ into the following intervals:

  \medskip
  \begin{center}
    \input{_fig3}
  \end{center}

  \medskip
  \noindent
  Consider the interval $I=[3,6]$.
  We find all occurrences of $U=bb \diamondsuit$ in $T$:

  \medskip
  \begin{center}
    \input{_fig4}
  \end{center}

  \medskip
  \noindent
  We have:
  $$E = \SolidOcc(U,T) = \{2,6,10,11,14\},\ H = \NonSolidOcc(U,T) = \{1,5,9,15,16\}.$$
  The positions in $E$ can be partitioned among two solid $\odot$-prefixes: $bba$ ($\{2,6,11\}$) and $bbb$ ($\{10,14\}$).
  For $bba$, all the three positions $j$ satisfy $\lcp(T[1..6],T[j..18])=3$ and each of them can be chosen as a representative.
  For $bbb$, the representative is at position 10 with $\lcp(T[1..6],T[10..18])=6$.
  
  We use the \ShortestCover$(S,L)$ subroutine for the following pairs $(S,L)$:
  $$(bba,\{1,2,5,6,9,11,15,16\})\ \mbox{ and }\ 
    (bbbabb,\{1,5,9,10,14,15,16\}).$$
  Only the latter call finds a cover: $bbbab$ with the covering set $\{1,5,10,14\}$:

  
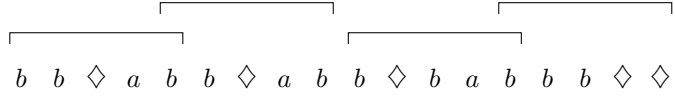
\medskip
  \begin{center}
    \input{_fig5}
  \end{center}

  \subsection{Covering with Non-$\odot$-prefixes}\label{ss:non-weak}

  In this section we are searching for the shortest cover of $T$ assuming that it is not a $\odot$-prefix.
  By Observation~\ref{obs:main}, such a cover $S$ may occur only at ambiguous positions.
  Moreover, it must admit a small covering set:
 
   \begin{lemma}\label{lem:scs}
   Let $S$ be a cover of $T$. If $S$ is not a $\odot$-prefix,
   then it has a covering set of size at most $2k$.
   \end{lemma}
   \begin{proof}
   Let $\C$ be a minimal covering set of $S$. Any factor $T[i..i+|S|-1]$ for $i\in\C$ is not solid,
   so it must cover a non-solid position of $T$. By Observation~\ref{obs:atmost2}, any position is covered by at most two
   such occurrences, so $|\C|\le 2k$.\mayqed
   \end{proof}

  For a set of positions $\P$,
  we introduce an auxiliary operation \TestCover$(\P)$ which checks
  there is a cover of $T$ for which $\P$ is a covering set.
  Note that the length of such a cover is fixed to $n+1-\max\P$.
  This operation is particularly simple to implement for partial words;
  see the following lemma.

  \begin{lemma}\label{lem:testcover}
    After $2^{\Oh(k)}+\Oh(nk^2)$-time preprocessing,
    \TestCover$(\P)$ can be implemented in $\Oh(|\P|k)$ time.
    If $T$ is a partial word, then $\Oh(nk^2)$-time preprocessing suffices.
  \end{lemma}
  \begin{proof}
    Let $m=n+1-\max\P$.
    First consider the simpler case when $T$ is a partial word.
    By definition, $\P$ can be a covering set for a cover of length $m$ if
    and only if $1 \in \P$ and $\maxgap(\P\cup \{n+1\})\le m$.
    These conditions can be easily checked in $\Oh(|\P|)$ time without any preprocessing.

    Now, it suffices to check if there is a solid string $S$ of length $m$
    such that $T[i..i+m-1] \approx S$ for all $i \in \P$.
    After $\Oh(nk^2)$-time preprocessing, we can compute $\lcp(1,i)$ for all $i \in \P$
    and check if each of those values is at least $m$.
    If not, then certainly such a string $S$ does not exist.
    Otherwise, let the set $Y$ contain positions of all don't care symbols in $T[1..m]$.
    We need to check, for each $j \in Y$, if the set
    $$X_j = \{T[i-1+j]\,:\,i \in \P\}$$
    contains no more than one solid symbol.
    This last step is performed in $\Oh(|\P|\, k)$ time.
    
    \medskip
    If $T$ is a general i-string, the only required change is related to processing the $X_j$ sets.
    If a set $X_j$ contains a solid symbol, then it suffices to check if this symbol
    matches all the other symbols in this set.
    Otherwise we need some additional preprocessing.

    Let $Z$ be the set of all non-solid positions in $T$.
    We wish to compute, for each subset of $Z$, if there is a single solid
    symbol matching all the positions in this subset.
    For this, we first reduce the size of the alphabet.
    For each solid symbol $c \in \Sigma$, we find the subset of $Z$ which contains this symbol.
    Note that if for two different solid symbols these subsets are equal, we can
    remove one of those symbols from the alphabet (just for the preprocessing phase).
    This way we reduce the alphabet size to at most $2^k$.
    Afterwards we simply consider each subset of $Z$ and look for a common solid symbol,
    which takes $2^{\Oh(k)}$ time.
  \mayqed\end{proof}

  \begin{theorem}\label{thm:first_fpt}
    The shortest cover of an i-string $T$ with $k$ non-solid symbols can be computed
    in $2^{\Oh(k \log k)} + \Oh(nk^4)$ time.
  \end{theorem}
  \begin{proof}
    By Theorem~\ref{thm:solid}, if the shortest cover of $T$ is a $\odot$-prefix then
    it can be computed in $\Oh(nk^4)$ time.
    Otherwise, by Lemma~\ref{lem:scs} such a cover  $S$
    has a minimal covering set of size at most $2k$. Moreover, 
    since $S$ may occur at ambiguous positions only, this covering set is a subset
    of $\A$. 
    We generate all subsets $\P\sub \A$ of size at most $2k$
    and for each of them run \TestCover$(\P)$. 
    The number of calls to \TestCover\ is
    $$\Oh\left(\sum_{i=1}^{2k}\tbinom{|\A|}{i}\right) = 
    \Oh\left(\sum_{i=1}^{2k}k^{2i}\right)= 2^{\Oh(k\log k)},$$
    and consequently the total running time of these calls, including preprocessing, is $\Oh(nk^2+k^22^{\Oh(k\log k)})=
    \Oh(nk^2)+2^{\Oh(k\log k)}$.
  \mayqed\end{proof}

  \subsection{More Efficient Algorithm for Partial Words}\label{subsec:partial}
 
  We conclude with an algorithm for partial words which is faster
  than the generic solution for i-strings.
  \begin{theorem}\label{thm:main}The shortest cover of a partial word of length $n$ with $k$ don't care symbols can be computed
    in $2^{\Oh(\sqrt{k} \log k)} + \Oh(nk^4)$ time.
  \end{theorem}
  \begin{proof}
    We improve the algorithm from the proof of Theorem~\ref{thm:first_fpt}.
    The only part of that algorithm that does not work in $\Oh(nk^4)$ time is
    searching for a cover under the assumption that it is not a $\odot$-prefix.
    Recall that such a cover $S$ may only occur at ambiguous positions.
    One of the occurrences must be a suffix of $T$, which restricts the length of such a cover to $n+1-i$ for $i\in \A$.
       Let us fix $m$ to be one of these lengths.
    
    Let $U=T[1..m]$ and let $\P \subseteq \A$ be the set of positions $i\in \A$ for which
    $U \odot i$ has at most $\sqrt{k}$ don't care symbols.
    We consider two cases.

    \medskip
    \noindent
    \textbf{Case 1:}
    $S$ has an occurrence $i \in \P$.
    Let $i_1,\ldots,i_r$ be the don't care positions in $U \odot i$.
    Let $M_1,\ldots,M_r$ be the sets of all solid symbols at positions $i_1,\ldots,i_r$
    in $U \odot j$ for $j \in \A$.
    If any of the sets $M_a$ is empty, we insert an arbitrary symbol from $\Sigma$ to it.

    Let us construct all possible solid strings by inserting symbols from $M_1,\ldots,M_r$
    at positions $i_1,\ldots,i_r$ in $U \odot i$.
    For each such solid string $S$, we simply compute a list $L$ of all positions $j \in \A$ such that
    $U \odot j \approx S$ and check if $1 \in L$ and if $\maxgap(L \cup \{n+1\})\le m$.
    Since $r \le \sqrt{k}$ and $|M_a| \le |\A| \le k^2$ for all $a=1,\ldots,r$, this shows that Case 1
    can be solved in $\Oh(k^{2\sqrt{k}+2}) = 2^{\Oh(\sqrt{k}\log{k})}$ time.

    \medskip
    \noindent
    \textbf{Case 2:} 
    $S$ has all its occurrences in $\A \setminus \P$.
    Let $\C \sub \A\sm \P$ be a minimal covering set of $S$.
    Note that each factor $T[i..i+|S|-1]$ for $i\in \C$ must contain
    at least $\sqrt{k}$ don't care symbols. By Observation~\ref{obs:atmost2},
    any don't care symbol can be covered by at most two such factors,
    which implies $|\C|\le 2\sqrt{k}$.
    We run \TestCover($\P$) for all sufficiently small subsets of $\A\setminus \P$.
    By Lemma~\ref{lem:testcover}, this requires 
    $2^{\Oh(\sqrt{k} \log k)} +\Oh(nk^2)$ time.
  \mayqed\end{proof}

  \section{Hardness Results}\label{sec:NP-hard}
  Negative results obtained for partial words remain valid in the more general setting of the i-strings,
  so in this section we restrict to partial words. We consider the following
  decision problem.

  \begin{problem}[\textsc{Shortest Cover in Partial Words}]\label{pr:decision-cover}
    Given a partial word $T$ of length~$n$ over an alphabet $\Sigma$ and an integer $d$,
    decide whether $T$ has a solid cover of length at most $d$.
  \end{problem}

  We devise a reduction from the CNF-SAT Problem.
  Recall that in this problem we are given a Boolean formula with $p$ variables which is a conjuntion of $m$ clauses
  $C_1 \land C_2 \land \ldots \land C_m$, where each clause $C_i$ is a disjunction
  of (positive or negative) literals, and our goal is to check if there exists an interpretation that satisfies
  the formula.
  Below we present a reformulation of the CNF-SAT Problem which is more suitable for our proof.

  \begin{problem}[\textsc{Universal Mismatch}]\label{pr:our-SAT}
    Given binary partial words $W_1,\ldots,W_m$ each of length $p$, check if there
    exists a binary partial word $V$ of length $p$ such that $V \not\approx W_i$ for any $i$.
  \end{problem}

  \begin{observation}\label{obs:refor}
   Given an instance of the CNF-SAT~Problem with $p$ variables and $m$ clauses,
   in linear time one can construct an equivalent instance of the \textsc{Universal Mismatch} Problem with $m$ partial words each of length $p$.
   The resulting mapping of instances is bijective and its inverse can also be computed in linear time.
  \end{observation}
  
  \begin{example}
    Consider a formula $\phi=(x_1 \lor x_2 \lor \neg x_3 \lor x_5) \land (\neg x_1 \lor x_4) \land (\neg x_2 \lor x_3 \lor \neg x_5)$
    with three clauses and five variables.
    In the corresponding instance of the \textsc{Universal Mismatch} Problem, for each clause $C_i$ we construct a partial
    word $W_i$ such that $W_i[j]=0$ if $x_j \in C_i$, $W_i[j]=1$ if $\neg x_j \in C_i$, and $W_i[j] = \diamondsuit$ otherwise:
    $$W_1 = 0 0 1 \diamondsuit 0, \quad W_2 = 1 \diamondsuit \diamondsuit 0 \diamondsuit, \quad W_3 = \diamondsuit 1 0 \diamondsuit 1.$$
    The interpretations $(1,0,1,1,0)$, $(1,1,1,1,0)$ satisfy $\phi$.
    They correspond to partial words $10110$, $11110$ and $1 \diamondsuit 110$, none of which
    matches any of the partial words $W_1$, $W_2$, $W_3$.
  \end{example}
  
  Consider an instance $\textbf{W}=(W_1,\ldots,W_m)$, $|W_j|=p$, of the \textsc{Universal Mismatch} Problem.
  We construct a binary partial word $T$ of length $\Oh(p(p+m))$ which is equivalent to $\textbf{W}$ as an instance of the \textsc{Shortest Cover in Partial Words} Problem with $d=4p+3$. 
  
  We define a morphism
  $$h:\quad 0 \rightarrow 0100,\quad 1 \rightarrow 0001,\quad \diamondsuit \rightarrow 0000,$$
  and construct $T$ so that a partial word $V$ of length $p$ is a solution to $\textbf{W}$ if and only if $S=11h(V)0$ covers~$T$.
  The word $T$ is of the form $11\pi^p0\beta_1\ldots\beta_p\gamma_{W_1}\ldots\gamma_{W_m}$,
  where $\pi=0\diamondsuit0\diamondsuit$ and $\beta_j$, $\gamma_{W}$ are \emph{gadgets} to be specified later.
  These gadgets are chosen so that every cover of $T$ has length at least $d$
  and every $d$-cover of $T$ (i.e., every cover of $T$ of length exactly $d$) is a $d$-cover of each gadget string $\beta_j$ and $\gamma_W$.
  Here, the prefix $11\pi^p0$ and all $\beta_j$ are \emph{consistency}
  gadgets which guarantee that any $d$-cover is of the form $11h(V)0$ for some partial word $V$ of length $p$.
  On the other hand, $\gamma_W$ are \emph{constraint} gadgets which do not allow $V$ to match $W$.

\subsection{Consistency Gadgets}  

  The prefix $11\pi^p0$ of $T$ enforces that
   any $d$-cover $S$ of $T$ is of the form $S=11s_1\ldots s_p0$  where $s_j\approx\pi$ for each~$j$. 
  Thus, in order to make sure that $S$ is of the form $11h(V)0$ for some partial word $V$,
  it suffices to rule out the possibility that $s_j=0101$ for some $j$.
  To this end, we define
  $$\beta_j = 11\, \pi^{p-1}\, 0\, \diamondsuit^{4j+1}\, 000\, \diamondsuit^d.$$

    \begin{figure}[ht]
      \centering
      \input{_fig_np_widgets_beta.tex}
      \caption{Sample gadget $\beta_j$ for $j=2$ and $p=3$ with occurrences of a pattern $11h(101)0$.
        Positions $d-2$ and $d$ are marked in grey.
      }
      \label{fig:np-widgets-beta}
    \end{figure}
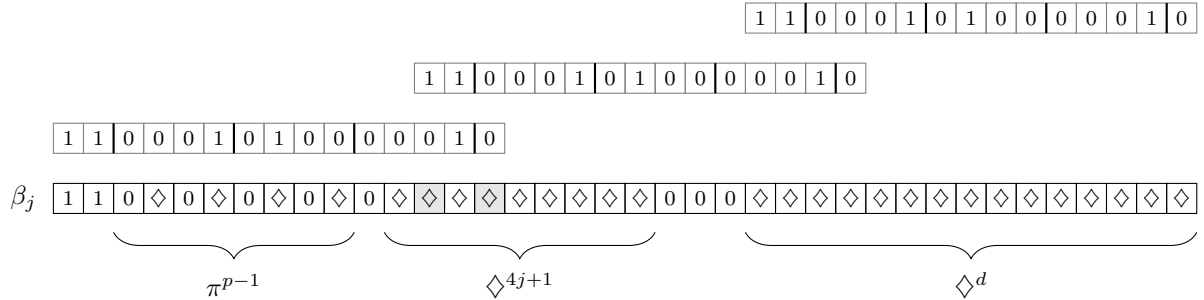
    
      \begin{observation}\label{obs:borb}
  Suppose $S$ is a solid string such that $S\approx 11\pi^p0$.
  Then $S$ occurs as a prefix and as a suffix of $\beta_j$.
  \end{observation}
  
  \begin{lemma}\label{lem:alpha_beta}
   Let $S=11s_1\ldots s_p0$ be a solid string with $s_i\approx \pi$ for each $i$. Then $S$ covers $\beta_j$
   if and only if $s_j\ne 0101$.
  \end{lemma}
  \begin{proof} 
    \noindent     $(\Leftarrow)$
      By Observation~\ref{obs:borb}, $S$ occurs in $\beta_j$ at positions $1$ and $|\beta_j|-|S|+1=d+4j+1$.
    If $s_j\ne 0101$, then $s_j=0100$ and $S$ also occurs at position $d-2$,
    or $s_j=0001$ and $S$ occurs at position $d$, or $s_j=0000$ and $S$ occurs at both positions $d-2$ and $d$; 
    see Figure~\ref{fig:np-widgets-beta}. Consequently, $S$ covers $\beta_j$
    since $\maxgap(1,d-2,d+4j+1)\le d$ and $\maxgap(1,d,d+4j+1)\le d$.

    $(\Rightarrow)$
    If $S$ covers $\beta_j$, it must have an occurrence at some position $q$ with $2\le q\le d+1$.
    In particular, $11$ must occur at position $q$, which further restricts $q \in \{d-3,d-2,d-1,d,d+1\}$.
    If $s_j=0101$, then we would need to have $\beta_j[q+4j-1]\approx 1$ and $\beta_j[q+4j+1]\approx 1$; see Figure~\ref{fig:np-widgets-beta}.
    However, $\beta_j[d+4j-2]=\beta_j[d+4j-1]=\beta_j[d+4j]=0$.
    We get a contradiction for each of the five possible values of $q$. Consequently, 
    $S$ cannot have $s_j=0101$.
  \mayqed\end{proof}
  \begin{corollary}\label{cor:alpha_beta}
  A solid string $S\approx 11\pi^p0$ is a cover of each partial word $\beta_j$ for $j=1,\ldots,p$ if and only if 
  $S=11h(V)0$ for  a binary partial word $V$ of length $p$.
  \end{corollary}

 \subsection{Constraint Gadgets}  

  We encode a constraint $V\not \approx W$ using a gadget
  $$\gamma_W=11\mu(W^R)010\diamondsuit^d$$
  where $W^R$ denotes the reverse of $W$
  and $\mu$ is the following morphism:
  $$
    \mu:\quad
    0 \rightarrow \diamondsuit\diamondsuit0\diamondsuit,
    \quad
    1 \rightarrow 0\diamondsuit\diamondsuit\diamondsuit,
    \quad
    \diamondsuit \rightarrow 0\diamondsuit0\diamondsuit.
  $$

  \begin{observation}\label{obs:borg}
  Suppose $S$ is a solid string such that $S\approx 11\pi^p0$ and $W$ is a partial word of length $p$.
  Then $S$ occurs as a prefix and as a suffix of $\gamma_W$.
  \end{observation}

  Before we proceed with a proof that $\gamma_W$ indeed encodes the constraint,
  let us characterize the relation between morphisms $\mu$ and $h$.

   \begin{lemma}\label{lem:h-mu}    Let $c, c'\in \{0,1,\diamondsuit\}$, and let $X$, $Y$ be partial words of the same length.
    Then $11h(Xc)0$ occurs in $\mu(c'Y)\,010\,\diamondsuit\diamondsuit$
    if and only if $c \not\approx c'$.
  \end{lemma}
           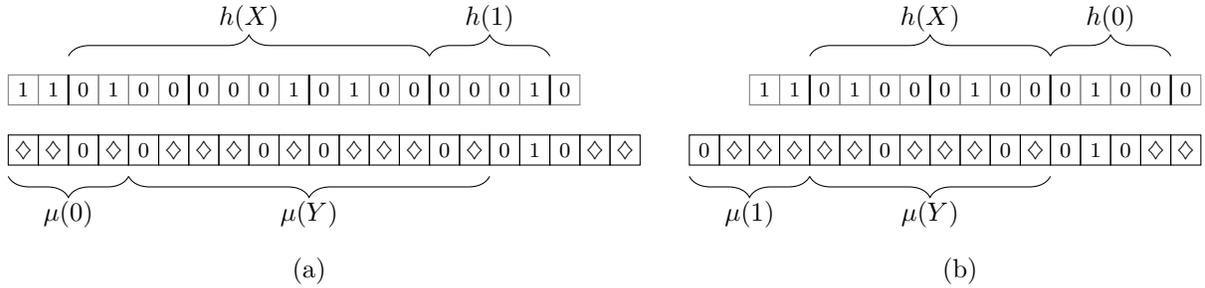
\begin{figure}[htpb]
	\begin{center}
    \input{_fig_np_widgets_lemma.tex}
    \end{center}
    \caption{
      Illustration of Lemma~\ref{lem:h-mu}:
      an occurrence of $11h(Xc)0$
      in $\mu(c'Y)\,010\,\diamondsuit\diamondsuit$ for
      (a) $Xc=0101$, $c'Y=01\diamondsuit 0$;
      (b) $Xc=000$, $c'Y=100$.
      In general, $11h(Xc)0$ is a prefix of  $\mu(c'Y)\,010\,\diamondsuit\diamondsuit$
      if $c=1$ and $c'=0$, and a suffix --- if $c=0$ and $c'=1$.
    }
    \label{fig:np-widgets-lemma}
  \end{figure}
  \begin{proof}
    Let $P=11h(Xc)0$, $Q=\mu(c'Y)010\diamondsuit\diamondsuit$ and $\ell=|P|$.
    
    ($\Rightarrow$) Note that $|Q|=\ell+2$, so $P$ can occur in $Q$ only at positions $p\in \{1,2,3\}$.
    Moreover, $p=2$ is impossible because  $Q[\ell-1]=1$ and $P[\ell-2]=0$ (since $h(c)\approx \pi= 0\diamondsuit0\diamondsuit$);
    see Figure~\ref{fig:np-widgets-lemma}.
    Thus, $P$ can occur in $Q$ only as a prefix or as a suffix.
    
    Suppose $P$ occurs as a prefix of $Q$.
    Note that $P$ begins with $11$, so $\mu(c')\approx 11\diamondsuit\diamondsuit$ and thus $c'=0$.
    Moreover, $Q$ ends with $010\diamondsuit\diamondsuit$, so $h(c)\approx \diamondsuit\diamondsuit01$ and $c=1$. 
    Similarly, if $P$ occurs as a suffix of $Q$, then $\mu(c')\approx \diamondsuit\diamondsuit11$, so $c'=1$,
    and $h(c)\approx 010\diamondsuit$, so $c=0$.  
    Consequently, $c\not\approx c'$ in either case.

    ($\Leftarrow$)
    Observe that $\mu(c'Y)$ has $\diamondsuit$'s at all even positions, and $\diamondsuit$'s or zeroes
    at all odd positions, while, $h(Xc)0$ has zeroes at all odd positions.
    Thus, any mismatch preventing an occurrence of $P$ as a prefix or as a suffix of $Q$
     must be due to the initial $11$ in $P$ or the terminal $010\diamondsuit\diamondsuit$ in $Q$.
    The corresponding positions in $Q$ and $P$ depend only on $c'$ and $c$, respectively.
    As $c \not\approx c'$, we have $c=1$ and $c'=0$ or $c=0$ and $c'=1$.
    In the former case $P$ occurs in $Q$ as a prefix, and in the latter it occurs as a suffix;
    see Figure~\ref{fig:np-widgets-lemma}.
  \mayqed\end{proof}

         \begin{figure}[h]
    \centering
    \input{_fig_np_widgets_gamma.tex}
    \caption{A gadget $\gamma\,_{\diamondsuit01}$ with occurrences of a pattern $11h(110)0$.
    }
    \label{fig:np-widgets-gamma}
  \end{figure}
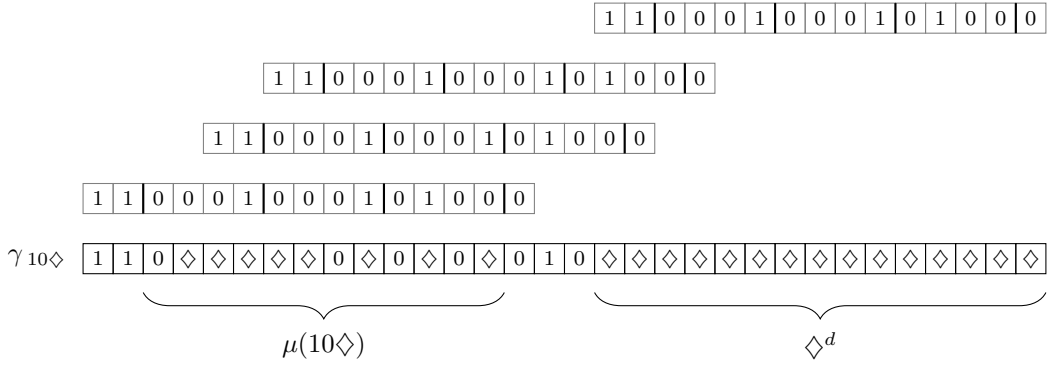
  
  \begin{lemma}\label{lem:gamma}
    Let $V$ and $W$ be binary partial words of length $p$.
    Then $S=11h(V)0$ covers $\gamma_W$ if and only if $V\not \approx W$.
  \end{lemma}
  \begin{proof}
  	 $(\Leftarrow)$ 
  	 Note that, by Observation~\ref{obs:borg}, $S$ always matches both a prefix and a suffix of $\gamma_W$.
    The only positions which are not covered by these two occurrences of $S$ form the middle $10$ factor $\gamma_{W}[d+1..d+2]$;
	see Figure~\ref{fig:np-widgets-gamma}.
    If $V\not \approx W$, there exists a position $i \in \{1,\ldots,p\}$
    such that $V[i]\not \approx W[i]$.
    By Lemma~\ref{lem:h-mu}, $11h(V[1..i])0$ occurs in $\mu((W[1..i])^R)010\diamondsuit\diamondsuit$.
    This occurrence extends to an occurrence of $11h(V)0$ in $\mu((W[1..i])^R)010\diamondsuit^{d-4i}$,
    and consequently an occurrence of $11h(V)0$ in $\gamma_W$ covering the middle $10$ factor $\gamma_W[d+1..d+2]$.
    Thus, $S=11h(V)0$ is a cover of $\gamma_W$.
    
    $(\Rightarrow)$
    Let $r$ be the position in $\gamma_{W}$ corresponding to an occurrence of $11h(V)0$
    that covers $\gamma_{W}[d+2]$. Note that $S$ begins with $11$, so $r<d-1$.
    Let $i=\lceil\frac{d-r}{4}\rceil$, i.e., $i$ is the smallest value such
    that the occurrence of $11h(V[1..i])0$ at position $r$ covers the middle $10$ factor $\gamma_W[d+1..d+2]$.
    Now, observe that $11h(V[1..i])0$ occurs in $\mu((W[1..i])^R)010\diamondsuit\diamondsuit$,
    so Lemma~\ref{lem:h-mu} implies that $V[i]\not \approx W[i]$,
    and thus $V\not\approx W$.
  \mayqed\end{proof}
  
\subsection{Main Negative Results}
  
  \begin{theorem}\label{thm:red}
  Given an instance $\textbf{W}$ of the \textsc{Universal Mismatch} Problem with $m$ partial words of length $p$,
  one compute in $\Oh(|T|)$ time a binary partial word $T$ of length $\Theta((p+m)^2)$
for which the \textsc{Shortest Cover in Partial Words} Problem with $d=4p+3$ is equivalent to $\textbf{W}$.
  \end{theorem}
  \begin{proof}
    Let  $$T=11\pi^p0\, \beta_1\ldots\beta_p\, \gamma\,_{W_1}\ldots\gamma\,_{W_m}.$$
    Each gadget $\beta_j$, $\gamma_W$ is of length $\Theta(p)$, so
    $|T|=\Theta((p+m)^2)$. Moreover, $T$ can clearly be constructed in $\Theta((p+m)^2)$ time.
    It suffices to prove that $\textbf{W}$ is a YES-instance of the \textsc{Universal Mismatch} Problem if and only if $(T,4p+3)$
    is a YES-instance of the \textsc{Shortest Cover in Partial Words} Problem.
    
    $(\Rightarrow)$ Suppose $\textbf{W}$ is a YES-instance with a solution $V$.
    We shall prove that a solid string $S=11h(V)0$ of length $d$ is a cover of $T$.
    We have $S\approx 11\pi^p0$ by definition of $h$ and $\pi$; in particular $S$ covers $11\pi^p0$.
    Moreover, $S$ covers each $\beta_j$ by Corollary~\ref{cor:alpha_beta},
    and for each $i$ it covers $\gamma_{W_i}$ by Lemma~\ref{lem:gamma} and due to the fact that $V\not\approx W_i$.
    Thus, $T$ is a concatenation of partial words covered by $S$, and thus $T$ itself is also covered by~$S$.

    $(\Leftarrow)$ Suppose that $T$ has a solid cover $S$ with $|S|\le d$.
   	Clearly, $|S|>1$ since both 0 and 1 occur as solid symbols in $T$.
   	Thus, $S$ begins with $11$. Note that $11$ does not occur in $T$ at any position $p$ with $1<p\le d$.
   	Consequently, $S$ cannot be shorter than $d$, i.e., $S\approx 11\pi^p0$.
   	
   	By Observations~\ref{obs:borb} and~\ref{obs:borg}, $S$ occurs both as a prefix and as a suffix of each gadget
   	words $\beta_j$ and $\gamma_W$.
   	It also covers their superstring $T$, so $S$ covers each of the gadget words.
   	By Corollary~\ref{cor:alpha_beta}, $S=11h(V)0$ for some partial word $V$,
   	and by Lemma~\ref{lem:gamma}, $V$ does not match any of the partial words $W_1,\ldots,W_m$.
  \end{proof}
  
  \begin{corollary}\label{cor:np-hard}
  The \textsc{Shortest Cover in Partial Words} Problem is NP-complete even for the binary alphabet.
  \end{corollary}
  \begin{proof}
    Equivalence between the \textsc{CNF-SAT} Problem and \textsc{Universal Mismatch} Problem (Observation~\ref{obs:refor})
  and the reduction above imply that the \textsc{Shortest Cover in Partial Words} Problem is NP-hard.
  It belongs to NP, since checking whether a given solid string is a cover can be implemented in polynomial time.
  \end{proof}

   The Exponential Time Hypothesis (ETH) \cite{DBLP:journals/jcss/ImpagliazzoP01,ETHsurvey} asserts that for some $\eps >0$
 the  3-CNF-SAT Problem cannot be solved in $\Oh(2^{\eps p})$ time, where $p$ is the number of variables.
 By the Sparsification Lemma~\cite{DBLP:journals/jcss/ImpagliazzoPZ01,ETHsurvey}, 
 ETH implies that for some $\eps>0$ the 3-CNF-SAT Problem cannot be solved in $\Oh(2^{\eps(p+m)})$ time,
 and consequently in $2^{o(p+m)}$ time, where $m$ is the number of clauses.
 Thus, Observation~\ref{obs:refor} and Theorem~\ref{thm:red} also imply the following result.
 
 \begin{corollary}\label{cor:lb}
     Unless the Exponential Time Hypothesis is false, there is no
    $2^{o(\sqrt{n})}$-time algorithm for the \textsc{Shortest Cover in Partial Words} Problem.
    In particular, there is no $2^{o(\sqrt{k})}n^{\Oh(1)}$-time algorithm for this problem.
 \end{corollary}

  \section{Conclusions}
  We considered the problems of finding the length of the shortest solid cover
  of an indeterminate string and of a partial word.
  The main results of the paper are fixed-parameter tractable algorithms for these problems parameterized by $k$,
  that is, the number of non-solid symbols in the input.
  For the partial word covering problem we obtain a $2^{\Oh(\sqrt{k}\log k)} + nk^{\Oh(1)}$-time algorithm
  whereas for covering a general indeterminate string we obtain a $2^{\Oh(k\log k)} + nk^{\Oh(1)}$-time algorithm.
  The latter can actually be improved to $2^{\Oh(k)} + nk^{\Oh(1)}$ time by extending the tools
  used in the proof of Theorem~\ref{thm:main}.
  In all our algorithms a shortest cover itself and all the lengths of covers could be computed
  without increasing the complexity.

  One open problem is to determine if the shortest cover of indeterminate strings
  can be found as fast as the shortest cover of partial words.
  Another question is to close the complexity gap for the latter problem,
  considering the lower bound resulting from the Exponential Time Hypothesis,
  which yields that no $2^{o(\sqrt{k})} n^{\Oh(1)}$-time solution exists for this problem.

\bibliographystyle{abbrv}
\bibliography{icover}

\end{document}

%% file: _fig1.tex
%
%

\begin{center}
\begin{tikzpicture}
\def\unitX{5mm}
\def\unitY{5mm}

\small

\foreach \x/\c in {1/b,2/b,3/\diamondsuit,4/\diamondsuit,5/a,6/b,7/b,8/\diamondsuit,9/\diamondsuit,10/b,11/a,12/\diamondsuit} {
    \draw node [above] at (\x*\unitX, 0) {$\bf \c$}; 
}

\foreach \x/\y/\label in {1/1/bbaa,2/3/bbaa,6/1/bbaa,9/3/bbaa} {
  \draw[yshift=0.2*\unitY] (\x*\unitX-0.3*\unitX,\y*\unitY*0.8+0.4*\unitY) -- (\x*\unitX-0.3*\unitX,\y*\unitY*0.8+0.7*\unitY)-- node[above] {\it \label} +(3.6*\unitX,0)--+(3.6*\unitX,-0.3*\unitY);
}

\end{tikzpicture}
\end{center}

%% file: _fig2.tex
%
%

\begin{center}
\begin{tikzpicture}
\def\unitX{5mm}
\def\unitY{5mm}

\small

\foreach \x/\c in {1/b,2/b,3/\diamondsuit,4/\diamondsuit,5/a,6/b,7/b,8/\diamondsuit,9/\diamondsuit,10/b,11/a,12/\diamondsuit} {
    \draw node [above] at (\x*\unitX, 0) {$\bf \c$}; 
}

\foreach \x/\y/\label in {1/1/bbab,3/3/bbab,6/1/bbab,9/3/bbab} {
  \draw[yshift=0.2*\unitY] (\x*\unitX-0.3*\unitX,\y*\unitY*0.8+0.4*\unitY) -- (\x*\unitX-0.3*\unitX,\y*\unitY*0.8+0.7*\unitY)-- node[above] {\it \label} +(3.6*\unitX,0)--+(3.6*\unitX,-0.3*\unitY);
}

\end{tikzpicture}
\end{center}

%% file: _fig3.tex
\begin{center}
\begin{tikzpicture}
\def\unitX{5mm}
\def\unitY{5mm}

\foreach \x/\c in {1/b,2/b,3/\diamondsuit,4/a,5/b,6/b,7/\diamondsuit,8/a,9/b, 10/b,11/\diamondsuit,12/b,13/a,14/b,15/b,16/b,17/\diamondsuit,18/\diamondsuit} {
    \draw node [above] at (\x*\unitX, 0) {$\c$}; 
}

\draw[snake=brace] (2.4*\unitX,-0.1*\unitY) -- node[below] {$[1,2]$} (0.6*\unitX,-0.1*\unitY);
\draw[snake=brace] (2.6*\unitX,1.1*\unitY)  -- node[above] {$[3,6]$} (6.4*\unitX,1.1*\unitY);
\draw[snake=brace] (10.4*\unitX,-0.1*\unitY) -- node[below] {$[7,10]$} (6.6*\unitX,-0.1*\unitY);
\draw[snake=brace] (10.6*\unitX,1.1*\unitY) -- node[above] {$[11,16]$} (16.4*\unitX,1.1*\unitY);
\draw[snake=brace] (17.4*\unitX,-.1*\unitY) -- node[below] {$[17,17]$} (16.6*\unitX,-.1*\unitY);
\draw[snake=brace]  (17.6*\unitX,1.1*\unitY) -- node[above] {$[18,18]$} (18.4*\unitX,1.1*\unitY);

\end{tikzpicture}
\end{center}

%% file: _fig4.tex
%
%

\begin{center}
\begin{tikzpicture}
\def\unitX{5mm}
\def\unitY{5mm}

\foreach \x/\c in {1/b,2/b,3/\diamondsuit,4/a,5/b,6/b,7/\diamondsuit,8/a,9/b, 10/b,11/\diamondsuit,12/b,13/a,14/b,15/b,16/b,17/\diamondsuit,18/\diamondsuit} {
    \draw node [above] at (\x*\unitX, 0) {$\c$}; 
}

\foreach \x/\y/\label in {1/1/bb\diamondsuit,2/3/bba,5/1/bb\diamondsuit,6/3/bba,9/1/bb\diamondsuit,10/3/bbb,11/5/bba,14/1/bbb,15/3/bb\diamondsuit,16/5/bb\diamondsuit} {
  \draw[yshift=0.2*\unitY] (\x*\unitX-0.3*\unitX,\y*\unitY*0.8+0.4*\unitY) -- (\x*\unitX-0.3*\unitX,\y*\unitY*0.8+0.7*\unitY)-- node[above] {$\label$} +(2.6*\unitX,0)--+(2.6*\unitX,-0.3*\unitY);
}

\end{tikzpicture}
\end{center}

%% file: _fig5.tex
%
%

\begin{center}
\begin{tikzpicture}
\def\unitX{5mm}
\def\unitY{5mm}

\foreach \x/\c in {1/b,2/b,3/\diamondsuit,4/a,5/b,6/b,7/\diamondsuit,8/a,9/b, 10/b,11/\diamondsuit,12/b,13/a,14/b,15/b,16/b,17/\diamondsuit,18/\diamondsuit} {
    \draw node [above] at (\x*\unitX, 0) {$\c$}; 
}

\foreach \x/\y/\label in {1/1/1,5/2/5,10/1/10,14/2/14} {
  \draw[yshift=0.2*\unitY] (\x*\unitX-0.3*\unitX,\y*\unitY*0.8+0.4*\unitY) -- (\x*\unitX-0.3*\unitX,\y*\unitY*0.8+0.7*\unitY)-- +(4.6*\unitX,0)--+(4.6*\unitX,-0.3*\unitY);
}

\end{tikzpicture}
\end{center}

%% file: _fig_np_widgets_beta.tex
\def\myfontsize{\footnotesize}
\newcommand{\occurrence}[3]{
    \foreach \c [count=\x from 0] in {#3} {
        \node at (#1*\boxWidth+\x*\boxWidth,#2*\boxWidth) {\myfontsize $\c$};
        \draw [gray, thin] (#1*\boxWidth+\x*\boxWidth-0.5*\boxWidth,#2*\boxWidth-0.5*\boxWidth)--+(0,\boxWidth)--+(\boxWidth,\boxWidth)--+(\boxWidth,0)--cycle;
    }
    \foreach \c [count=\x from 0] in {#3} {
        \pgfmathparse{Mod(\x,4)==2?1:0}
        \ifnum\pgfmathresult>0
            \draw [black, thick] (#1*\boxWidth+\x*\boxWidth-0.5*\boxWidth,#2*\boxWidth-0.5*\boxWidth)--+(0,\boxWidth);
        \fi
    }
}

\def\boxWidth{0.8}

\newcommand{\widget}[2]{
    \foreach \c [count=\x from 0] in {#2} {
        \node at (\x*\boxWidth,0) {\myfontsize $\c$};
        \draw (\x*\boxWidth-0.5*\boxWidth,-0.5*\boxWidth)--+(0,\boxWidth)--+(\boxWidth,\boxWidth)--+(\boxWidth,0)--cycle;
    }
    \node at (-0.75*\boxWidth,0) [left] {#1};
}

\begin{tikzpicture}[scale=0.5]
    \begin{scope}[yshift=-6cm]
        \occurrence{0}{2}{1,1,0,0,0,1,0,1,0,0,0,0,0,1,0}
        \occurrence{12}{4}{1,1,0,0,0,1,0,1,0,0,0,0,0,1,0}
        \occurrence{23}{6}{1,1,0,0,0,1,0,1,0,0,0,0,0,1,0}
        \filldraw[white!90!black] (12*\boxWidth-0.5*\boxWidth,-0.5*\boxWidth) rectangle (13*\boxWidth-0.5*\boxWidth,0.5*\boxWidth);
        \filldraw[white!90!black] (14*\boxWidth-0.5*\boxWidth,-0.5*\boxWidth) rectangle (15*\boxWidth-0.5*\boxWidth,0.5*\boxWidth);
        \widget{$\beta_j$}{
            1,1,
            0,\diamondsuit,0,\diamondsuit,
            0,\diamondsuit,0,\diamondsuit,
            0,\diamondsuit,
            \diamondsuit,\diamondsuit,\diamondsuit,\diamondsuit,
            \diamondsuit,\diamondsuit,\diamondsuit,\diamondsuit,
            0,0,0,
            \diamondsuit,\diamondsuit,\diamondsuit,\diamondsuit,
            \diamondsuit,\diamondsuit,\diamondsuit,\diamondsuit,
            \diamondsuit,\diamondsuit,\diamondsuit,\diamondsuit,
            \diamondsuit,\diamondsuit,\diamondsuit
        }
        \draw [decorate,decoration={brace,amplitude=10pt},xshift=0pt,yshift=-0.5cm]
          (9.5*\boxWidth,-0.5*\boxWidth) -- (1.5*\boxWidth,-0.5*\boxWidth)
          node [black,midway,yshift=-0.7cm] {$\pi^{p-1}$};
        \draw [decorate,decoration={brace,amplitude=10pt},xshift=0pt,yshift=-0.5cm]
          (19.5*\boxWidth,-0.5*\boxWidth) -- (10.5*\boxWidth,-0.5*\boxWidth)
          node [black,midway,yshift=-0.7cm] {$\diamondsuit^{4j+1}$};
        \draw [decorate,decoration={brace,amplitude=10pt},xshift=0pt,yshift=-0.5cm]
          (37.5*\boxWidth,-0.5*\boxWidth) -- (22.5*\boxWidth,-0.5*\boxWidth)
          node [black,midway,yshift=-0.7cm] {$\diamondsuit^d$};
    \end{scope}

\end{tikzpicture}

%% file: _fig_np_widgets_lemma.tex
\def\myfontsize{\footnotesize}
\newcommand{\occurrence}[3]{
    \foreach \c [count=\x from 0] in {#3} {
        \node at (#1*\boxWidth+\x*\boxWidth,#2*\boxWidth) {\myfontsize $\c$};
        \draw [gray, thin] (#1*\boxWidth+\x*\boxWidth-0.5*\boxWidth,#2*\boxWidth-0.5*\boxWidth)--+(0,\boxWidth)--+(\boxWidth,\boxWidth)--+(\boxWidth,0)--cycle;
    }
    \foreach \c [count=\x from 0] in {#3} {
        \pgfmathparse{Mod(\x,4)==2?1:0}
        \ifnum\pgfmathresult>0
            \draw [black, thick] (#1*\boxWidth+\x*\boxWidth-0.5*\boxWidth,#2*\boxWidth-0.5*\boxWidth)--+(0,\boxWidth);
        \fi
    }
}

\def\boxWidth{0.8}

\newcommand{\widget}[2]{
    \foreach \c [count=\x from 0] in {#2} {
        \node at (\x*\boxWidth,0) {\myfontsize $\c$};
        \draw (\x*\boxWidth-0.5*\boxWidth,-0.5*\boxWidth)--+(0,\boxWidth)--+(\boxWidth,\boxWidth)--+(\boxWidth,0)--cycle;
    }
    \node at (-0.75*\boxWidth,0) [left] {#1};
}

\twocol{
\begin{center}
\begin{tikzpicture}[scale=0.5]
    \begin{scope}
        \occurrence{0}{2}{1,1,0,1,0,0,0,0,0,1,0,1,0,0,0,0,0,1,0}
        \widget{}{
          \diamondsuit,\diamondsuit,0,\diamondsuit,
          0,\diamondsuit,\diamondsuit,\diamondsuit,
          0,\diamondsuit,0,\diamondsuit,
          \diamondsuit,\diamondsuit,0,\diamondsuit,
          0,1,0,\diamondsuit,\diamondsuit
        }
        \draw [decorate,decoration={brace,amplitude=8pt},xshift=0pt,yshift=-0.3cm]
          (3.5*\boxWidth,-0.5*\boxWidth) -- (-0.5*\boxWidth,-0.5*\boxWidth)
          node [black,midway,yshift=-0.5cm] {$\mu(0)$};

        \draw [decorate,decoration={brace,amplitude=8pt},xshift=0pt,yshift=-0.3cm]
          (15.5*\boxWidth,-0.5*\boxWidth) -- (3.5*\boxWidth,-0.5*\boxWidth)
          node [black,midway,yshift=-0.5cm] {$\mu(Y)$};

        \draw [decorate,decoration={brace,amplitude=8pt},xshift=0pt,yshift=-0.3cm]
          (1.5*\boxWidth,3.5*\boxWidth) -- (13.5*\boxWidth,3.5*\boxWidth)
          node [black,midway,yshift=0.5cm] {$h(X)$};

        \draw [decorate,decoration={brace,amplitude=8pt},xshift=0pt,yshift=-0.3cm]
          (13.5*\boxWidth,3.5*\boxWidth) -- (17.5*\boxWidth,3.5*\boxWidth)
          node [black,midway,yshift=0.5cm] {$h(1)$};

        \draw (9.5*\boxWidth,-4*\boxWidth) node {(a)};

    \end{scope}
  \end{tikzpicture}
  \end{center}
}{
\begin{center}
\begin{tikzpicture}[scale=0.5]
    \begin{scope}
        \occurrence{2}{2}{1,1,0,1,0,0,0,1,0,0,0,1,0,0,0}
        \widget{}{
          0,\diamondsuit,\diamondsuit,\diamondsuit,
          \diamondsuit,\diamondsuit,0,\diamondsuit,
          \diamondsuit,\diamondsuit,0,\diamondsuit,
          0,1,0,\diamondsuit,\diamondsuit
        }
        \draw [decorate,decoration={brace,amplitude=8pt},xshift=0pt,yshift=-0.3cm]
          (3.5*\boxWidth,-0.5*\boxWidth) -- (-0.5*\boxWidth,-0.5*\boxWidth)
          node [black,midway,yshift=-0.5cm] {$\mu(1)$};

        \draw [decorate,decoration={brace,amplitude=8pt},xshift=0pt,yshift=-0.3cm]
          (11.5*\boxWidth,-0.5*\boxWidth) -- (3.5*\boxWidth,-0.5*\boxWidth)
          node [black,midway,yshift=-0.5cm] {$\mu(Y)$};

        \draw [decorate,decoration={brace,amplitude=8pt},xshift=0pt,yshift=-0.3cm]
          (3.5*\boxWidth,3.5*\boxWidth) -- (11.5*\boxWidth,3.5*\boxWidth)
          node [black,midway,yshift=0.5cm] {$h(X)$};

        \draw [decorate,decoration={brace,amplitude=8pt},xshift=0pt,yshift=-0.3cm]
          (11.5*\boxWidth,3.5*\boxWidth) -- (15.5*\boxWidth,3.5*\boxWidth)
          node [black,midway,yshift=0.5cm] {$h(0)$};

        \draw (8.5*\boxWidth,-4*\boxWidth) node {(b)};

    \end{scope}
\end{tikzpicture}
\end{center}
}

%% file: _fig_np_widgets_gamma.tex
\def\myfontsize{\footnotesize}
\newcommand{\occurrence}[3]{
    \foreach \c [count=\x from 0] in {#3} {
        \node at (#1*\boxWidth+\x*\boxWidth,#2*\boxWidth) {\myfontsize $\c$};
        \draw [gray, thin] (#1*\boxWidth+\x*\boxWidth-0.5*\boxWidth,#2*\boxWidth-0.5*\boxWidth)--+(0,\boxWidth)--+(\boxWidth,\boxWidth)--+(\boxWidth,0)--cycle;
    }
    \foreach \c [count=\x from 0] in {#3} {
        \pgfmathparse{Mod(\x,4)==2?1:0}
        \ifnum\pgfmathresult>0
            \draw [black, thick] (#1*\boxWidth+\x*\boxWidth-0.5*\boxWidth,#2*\boxWidth-0.5*\boxWidth)--+(0,\boxWidth);
        \fi
    }
}

\def\boxWidth{0.8}

\newcommand{\widget}[2]{
    \foreach \c [count=\x from 0] in {#2} {
        \node at (\x*\boxWidth,0) {\myfontsize $\c$};
        \draw (\x*\boxWidth-0.5*\boxWidth,-0.5*\boxWidth)--+(0,\boxWidth)--+(\boxWidth,\boxWidth)--+(\boxWidth,0)--cycle;
    }
    \node at (-0.75*\boxWidth,0) [left] {#1};
}

\begin{tikzpicture}[scale=0.5]
    \begin{scope}[yshift=-15cm]
        \occurrence{0}{2}{1,1,0,0,0,1,0,0,0,1,0,1,0,0,0}
        \occurrence{4}{4}{1,1,0,0,0,1,0,0,0,1,0,1,0,0,0}
        \occurrence{6}{6}{1,1,0,0,0,1,0,0,0,1,0,1,0,0,0}
        \occurrence{17}{8}{1,1,0,0,0,1,0,0,0,1,0,1,0,0,0}
        \widget{$\gamma\,_{10\diamondsuit}$}{
            1,1,
            0,\diamondsuit,\diamondsuit,\diamondsuit,
            \diamondsuit,\diamondsuit,0,\diamondsuit,
            0,\diamondsuit,0,\diamondsuit,
            0,1,0,
            \diamondsuit,\diamondsuit,\diamondsuit,\diamondsuit,
            \diamondsuit,\diamondsuit,\diamondsuit,\diamondsuit,
            \diamondsuit,\diamondsuit,\diamondsuit,\diamondsuit,
            \diamondsuit,\diamondsuit,\diamondsuit
        }
        \draw [decorate,decoration={brace,amplitude=10pt},xshift=0pt,yshift=-0.5cm]
          (13.5*\boxWidth,-0.5*\boxWidth) -- (1.5*\boxWidth,-0.5*\boxWidth)
          node [black,midway,yshift=-0.7cm] {$\mu(10\diamondsuit)$};
        \draw [decorate,decoration={brace,amplitude=10pt},xshift=0pt,yshift=-0.5cm]
          (31.5*\boxWidth,-0.5*\boxWidth) -- (16.5*\boxWidth,-0.5*\boxWidth)
          node [black,midway,yshift=-0.7cm] {$\diamondsuit^d$};
    \end{scope}

\end{tikzpicture}